\documentclass{article}

\usepackage{graphicx}				
\usepackage{amsmath}				
\usepackage{amsfonts}
\usepackage[colorlinks=true,linkcolor=blue,anchorcolor=black,citecolor=blue,filecolor=black,menucolor=black,urlcolor=blue,breaklinks=true,pdfhighlight=/P,pdfmenubar=true,pdftoolbar=true,pdfpagelabels=true,pdfstartpage=1,pdfstartview=FitV,pdftitle={Pedestrian Route Choice by Iterated Equilibrium Search},pdfsubject={Pedestrian, Crowd, Mass, Simulation, Dynamics, VISWALK, VISSIM},pdfauthor={Kretz},pdfcreator={tobias Kretz},pdfproducer={t kretz},pdfkeywords={Pedestrian, Crowd, Mass, Simulation, Dynamics, VISWALK}]{hyperref}	
\usepackage[numbers,sort&compress]{natbib}	
\usepackage{hypernat}				
\usepackage[american]{babel}		
\usepackage{simplemargins}
\setallmargins{1in}
\usepackage[american]{babel}		
\title{Pedestrian Route Choice by Iterated Equilibrium Search}

\author{Tobias Kretz (corresponding author)$^1$, Karsten Lehmann$^{1,2}$, Ingmar Hofs{\"a}{\ss}$^1$ \\
$^1$: PTV Group, Haid-und-Neu-Stra{\ss}e 15, D-76131 Karlsruhe, Germany\\
\tt{\{Firstname.Lastname\}@ptvgroup.com} \\
phone: +49 721 9651 7280 \\
$^2$: init AG, K{\"a}ppelestra{\ss}e 4-6, D-76131 Karlsruhe, Germany
}%

\begin{document}
\maketitle

\abstract{In vehicular traffic planning it is a long standing problem how to assign demand such on the available model of a road network that an equilibrium with regard to travel time or generalized costs is realized. For pedestrian traffic this question can be asked as well. However, as the infrastructure of pedestrian dynamics is not a network (a graph), but two-dimensional, there is in principle an infinitely large set of routes. As a consequence none of the iterating assignment methods developed for road traffic can be applied for pedestrians. In this contribution a method to overcome this problem is briefly summarized and applied with an example geometry which as a result is enhanced with routes with intermediate destination areas of certain shape. The enhanced geometry is used in some exemplary assignment calculations.}

\section{Introduction}
For pedestrians (and vehicles alike) it holds that in general travel times along a route increase with increasing demand. This is because walking speeds are lower when density is higher and because higher demand implies larger queues in front of bottlenecks and therefore higher (enforced) dwell times. Obviously when there is more than one route available which connects the same origin and destination it can make sense -- with regard to the objective of minimizing travel times -- that a fraction of the pedestrians walks along these alternative routes. The process to calculate these fractions such that either all routes have the same travel time (user equilibrium) or the average of travel times is minimal (system optimum) is called (dynamic) assignment.

For vehicular traffic research on assignment methods has a history of more than a half century \cite{wardrop1952some,Beckmann1956,leblanc1975efficient,bar2002origin,gentile2009linear} and applying these methods is an established and major aspect of traffic planning.

Concerning pedestrians we have to go one step back and ask: what is a route? Or better: what distinguishes two different routes? It does not make sense to say -- with regard to assignment methods -- that two pedestrians have been walking two different routes, if their paths are the same everywhere except for a few millimeters. Instead we require that two paths only then belong to two different routes, if they cannot be transformed continuously one into the other without moving a path during the transformation over an obstacle of a given minimum extent. Recently we have proposed such a method \cite{Kretz2014h,Kretz2013b,Kretz2014a} and will apply and demonstrate it with an example geometry and in combination with a simple assignment method in the remainder of this contribution.

Before we do so, one has to ask first for the motivation or better for the benefit. The benefit of a method to find a set of relevant and mutually sufficiently distinct routes in a pedestrian walking geometry probably is evident to every expert in the field. However, is there a benefit to compute a user equilibrium for pedestrian dynamics? After all there is not such a eminent day-by-day commuting traffic as it is in vehicular traffic and day-by-day experience is {\em the} argument which explains why in road traffic there should be an equilibrium. Yet there are comparable situations in pedestrian traffic. Public transport commuters meet daily in large numbers in large stations. Their day-by-day experience may suggest them to use a route which may be longer or more arduous but saves them time; compare for example in this book \cite{Heuvel2014} where the average travel times of commuters on an escalator and those walking stairs (not parallel to the escalators) are astonishingly similar. A fire safety engineer can have a totally different motivation to compute a user equilibrium: s/he may not assume that the user equilibrium emerges spontaneously from the system if every occupant decides on his or her own which route to take. But knowing that the user equilibrium usually also is an efficient distribution on the available routes with relatively small delay and travel times the fire safety engineer might want to take the results from the assignment calculation as basis for the escape plan, i.e. as answer to the question where an escape route sign has to point into one direction and where into another one.

At the end of this introduction we would like to point the reader to another contribution in this book which approaches the pedestrian assignment with a different method \cite{vanwageningenkessels2014}.

\section{Example Application}
\begin{figure}[htbp]%
\center
\includegraphics[width=0.612\columnwidth]{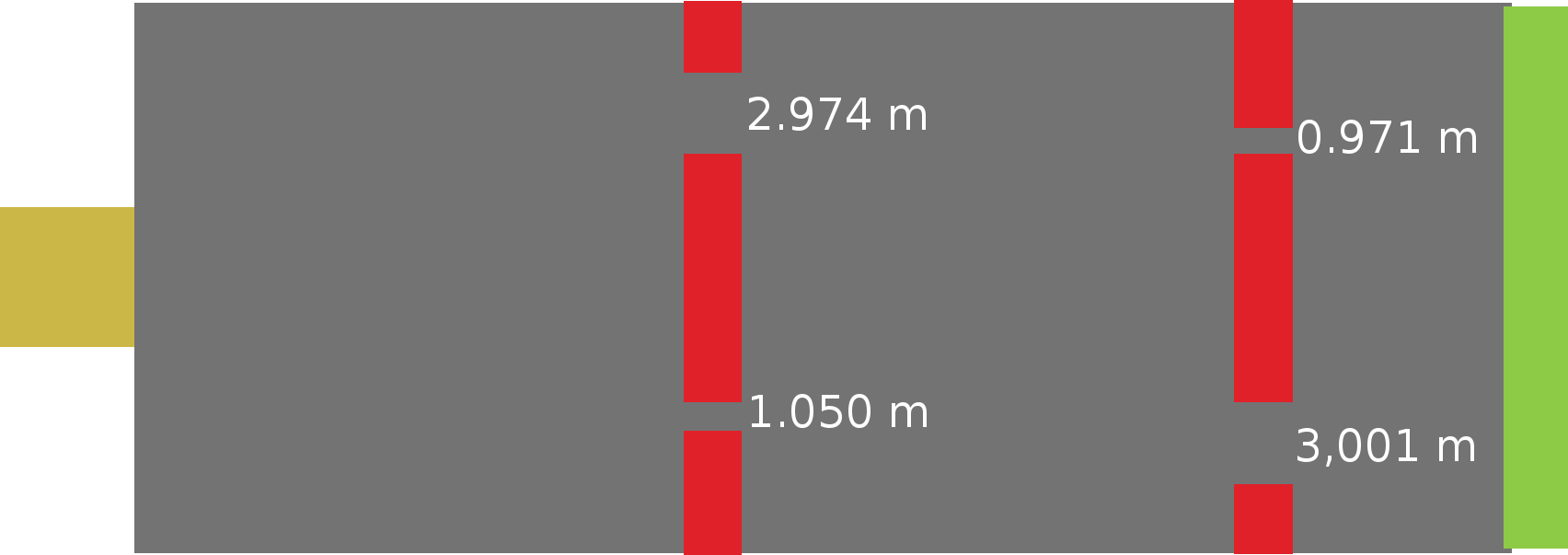}%
\caption{Walking geometry. Pedestrians are set with a given rate and spatially equally distributed into the simulation on the yellow area to the left, have to pass through the bottlenecks formed by the red obstacles and arrive on the green area on the right side.}%
\label{fig:Tuerbreiten}%
\end{figure}

\begin{figure}[htbp]%
\center
\includegraphics[width=0.612\columnwidth]{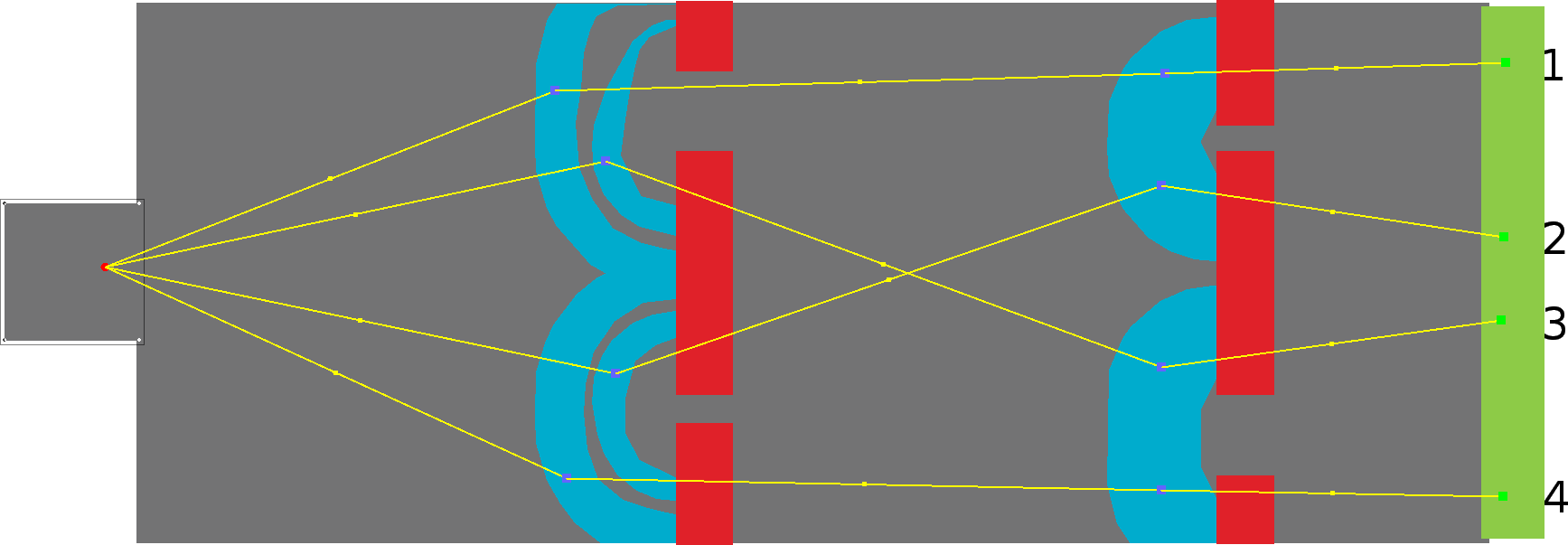}%
\caption{Walking geometry enhanced with automatically computed routes (yellow) and their intermediate destinations (blue). Note the numbers by which the routes will be identified in the text.}%
\label{fig:Routen}%
\end{figure}

Figure \ref{fig:Tuerbreiten} shows a walking geometry where pedestrians have to pass two times through one of two differently wide doors. Figure \ref{fig:Routen} shows the same geometry enhanced by routes and their intermediate destinations which have been calculated automatically with the method introduced in \cite{Kretz2014h}. 

For this work operationally pedestrians are simulated using PTV Viswalk \cite{Viswalk540,Kretz2008b,fellendorf2010microscopic} which itself utilizes a combination of the circular specification and the elliptical specification II of the Social Force Model \cite{Johansson2007}. The pedestrians are approaching the intermediate destination with the direction of the desired velocity set into the direction of the spatially shortest path to the next intermediate destination area. This implies that pedestrians will arrive at an intermediate destination usually walking orthogonally to the boundary of its area. Any deviation from this must be due to forces between pedestrians. As the upstream borders of all intermediate destinations are computed such that each point on that border has the same distance to the closest point on the next downstream intermediate destination this implies that pedestrians will not at all make a turn when they have reached an intermediate destination area and then proceed to the next one.

On the tactical level the distribution of the pedestrians on the four routes is done iteratively by he assignment method. In each iteration step one simulation is carried out. In the first iteration step 25\% of the pedestrians are sent on each of the routes. For each pedestrian the travel time is measured from the point where the pedestrian leaves the origin area to the moment when he reaches the destination area. From the average travel times on each of the routes the route choice ratios for the next iteration step are computed according to the following equation.

It is easy to see that an equilibrium distribution on the four available routes must depend on the total pedestrian volume set into the simulation (demand volume). We have done the iterated assignment with eleven different demand volumes: 0.5, 1.0, 1.5, 2.0, 2.5, 3.0, 3.5, 4.0, 4.5, 5.0, and 6.0 pedestrians per second. These numbers cover the whole range from total demand being below the capacity of the more narrow doors and total demand exceeding global capacity. For each demand volume the assignment procedure was carried out five times with different seed values for the random number generator. 

As termination condition was chosen that the largest and the smallest average travel time must not differ by more than 0.5 seconds. This is a relatively strict condition given the variation of walking speeds (default for men and women of an age of 30 to 50 years as required by the International Maritime Organization \cite{msc1238}). For this example study it was chosen this way to guarantee to see the dynamics of the assignment process unfold to the end. This comes at the danger of running into oscillations of a cycle length of two to four or five iteration steps toward the end of the assignment process when equilibrium is in principle reached, but the stochastic fluctuations are large enough that the termination condition is always missed by a small amount. In such a case we chose that iteration as result which came closest to the termination condition. 

In the assignment process in each iteration step (a step is one simulation run) the travel times of all pedestrians who arrived at the destination area between $t=300$ and $t=600$ seconds after the beginning of the simulation were recorded. At the end of the simulation the average travel time on each route was calculated. Then for the next iteration steps the route choice probability for the route with the smallest average travel time $t_{min}$ was increased and the route choice probability for the route with the largest average travel time $t_{max}$ was decreased by the same amount. This probability shift $\Delta p$ was calculated as

\begin{equation}
\Delta p = \alpha \left(\frac{t_{max} - t_{min}}{t_{max} + t_{min}}\right)^\delta
\end{equation}
where $\alpha$ is a general sensitivity factor which was chosen to be $\alpha=0.1$ in all computations and $\delta$ is a dynamic adaptation factor which usually was $\delta=1$, but was decreased when the routes with the longest and the smallest travel time were identical in subsequent iterations and which was increased when they exchanged roles in subsequent iterations.

\begin{figure}[htbp]%
\center
\includegraphics[width=0.612\columnwidth]{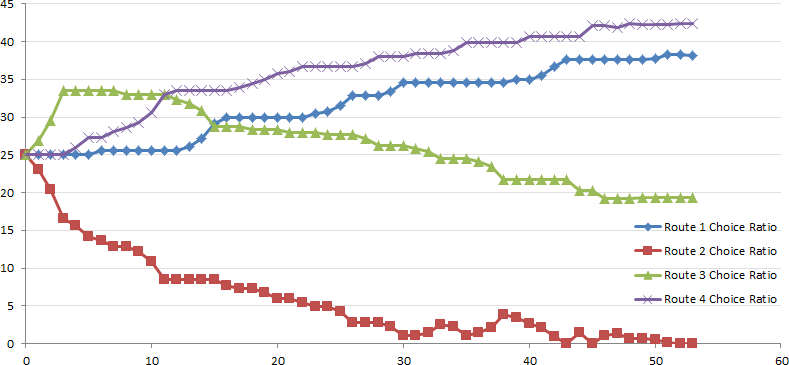} \\ \vspace{12pt}
\includegraphics[width=0.612\columnwidth]{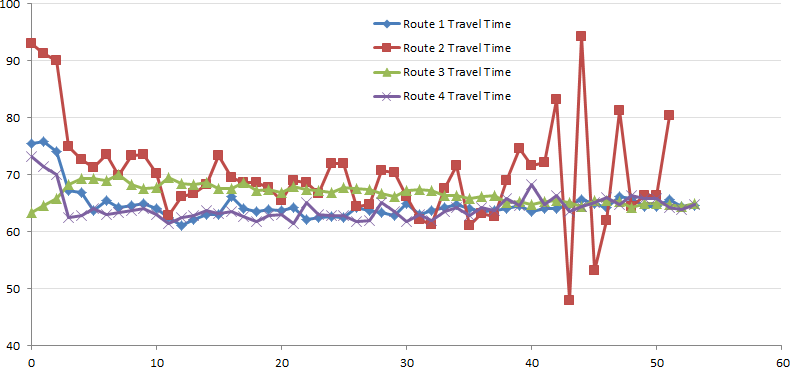}
\caption{Route choice ratios (top, in percent) and travel times (bottom, in seconds) for all four routes in the course of iterations if demand is 3 pedestrians per second. The travel time for route 2 becomes unstable when its route choice ratio is small (below 5\%) as then only few pedestrians (20 or even less) walk along route 2 and thus the average travel time is based only on few values.}%
\label{fig:results1}%
\end{figure}

Figure \ref{fig:results1} shows for a demand of 3 persons per second how route choice ratios and average travel times on each of the four routes evolve in the iteration process.

\begin{figure}[htbp]%
\center
\includegraphics[width=0.612\columnwidth]{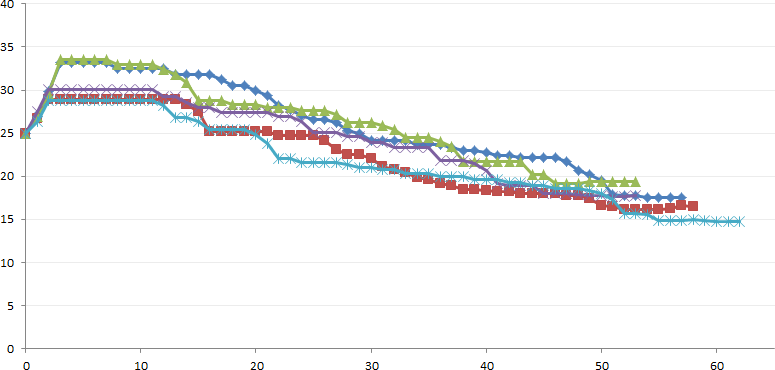} \\ \vspace{12pt}
\includegraphics[width=0.612\columnwidth]{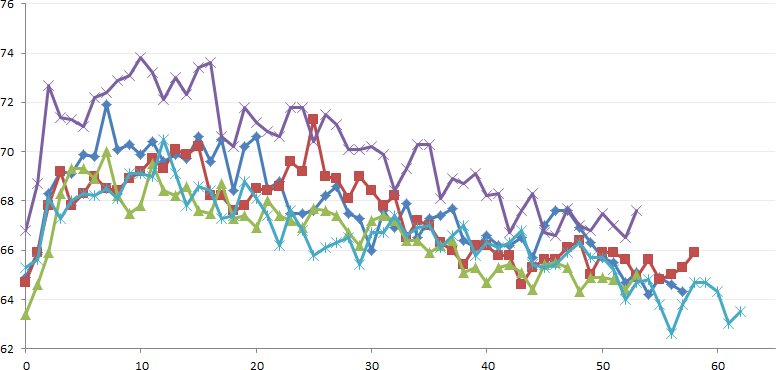}
\caption{Route choice ratios (top, in percent) and travel times (bottom, in seconds) for route 3 if demand is 3 pedestrians per second for all five iteration processes in comparison. While a general trend appears similar in all five processes there are differences which at least partly have their cause in the wide distribution of desired walking speeds of the pedestrians.}%
\label{fig:results2}%
\end{figure}

As described above we have carried out five assignment processes with different random numbers. It is of course interesting to compare how different different assignment processes evolve. For the sake of clarity we do not force travel times and route choice ratios of all routes into a diagram, but compare in figure \ref{fig:results2} the route choice ratios and travel times of the five different assignment processes only for route 3.

\begin{figure}[htbp]%
\center
\includegraphics[width=0.45\columnwidth]{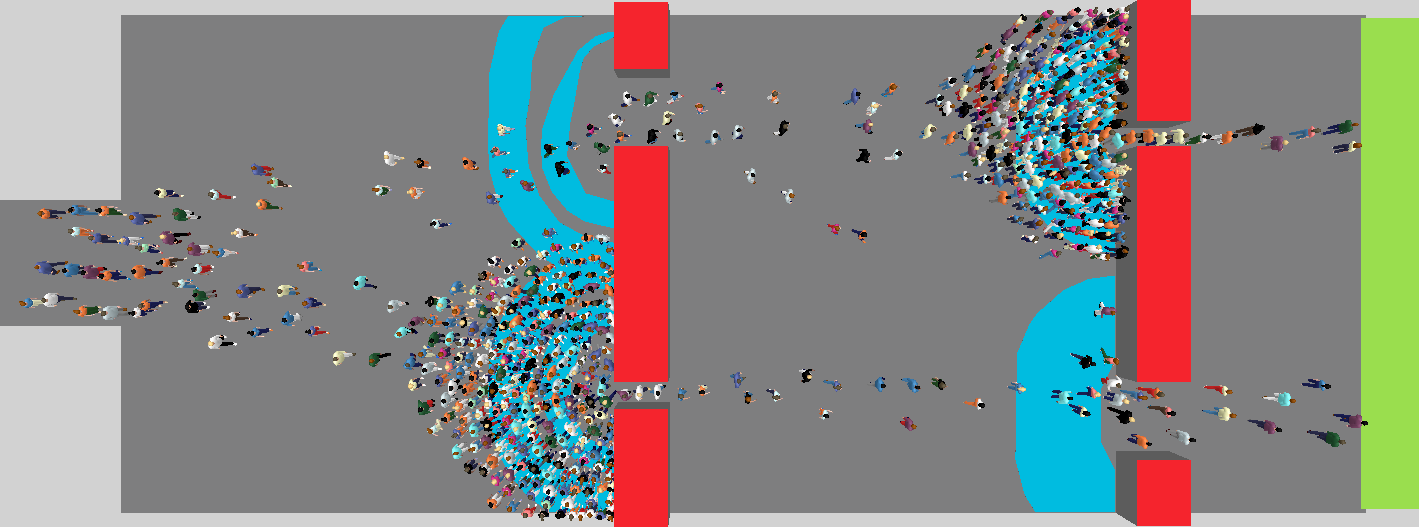} \hspace{12pt} 
\includegraphics[width=0.45\columnwidth]{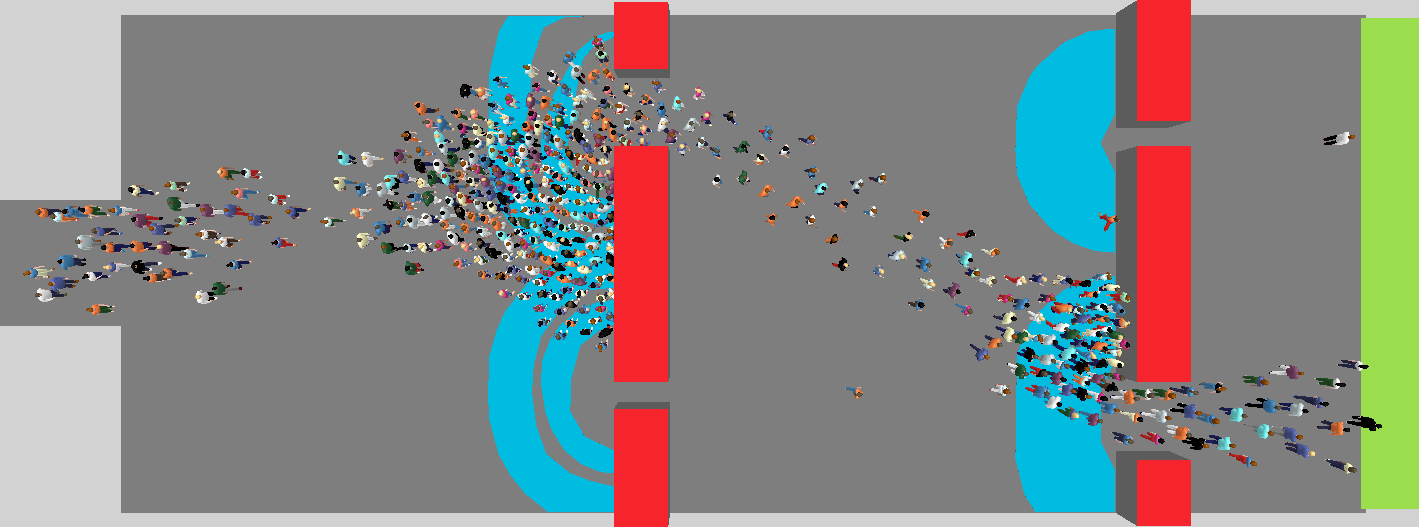} \\ \vspace{12pt}
\includegraphics[width=0.45\columnwidth]{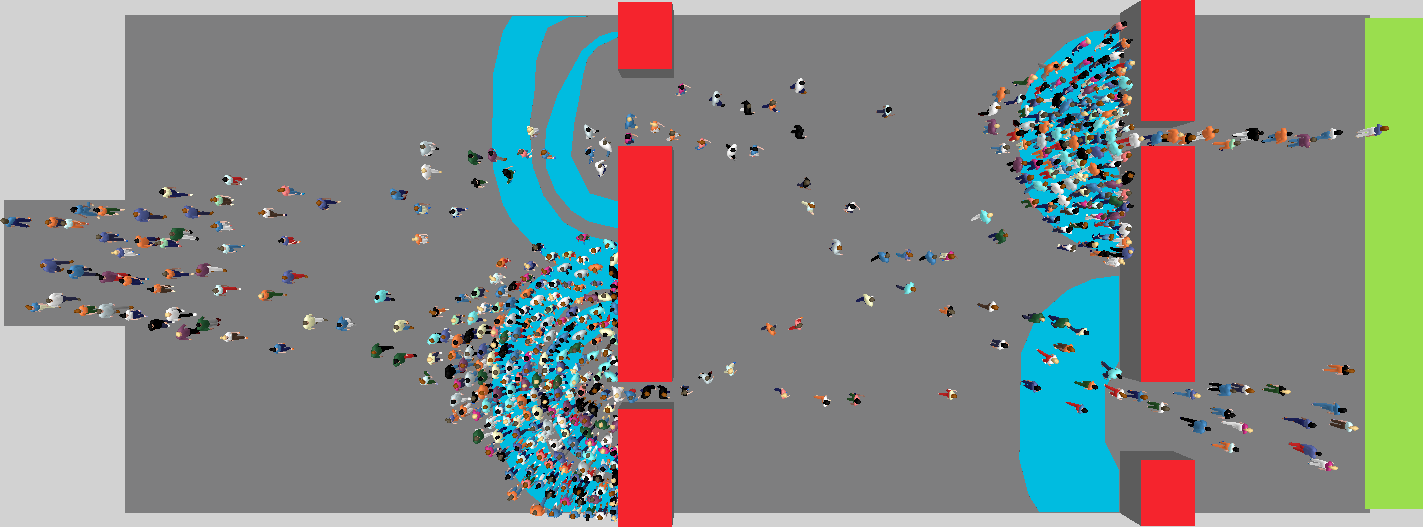} \hspace{12pt} 
\includegraphics[width=0.45\columnwidth]{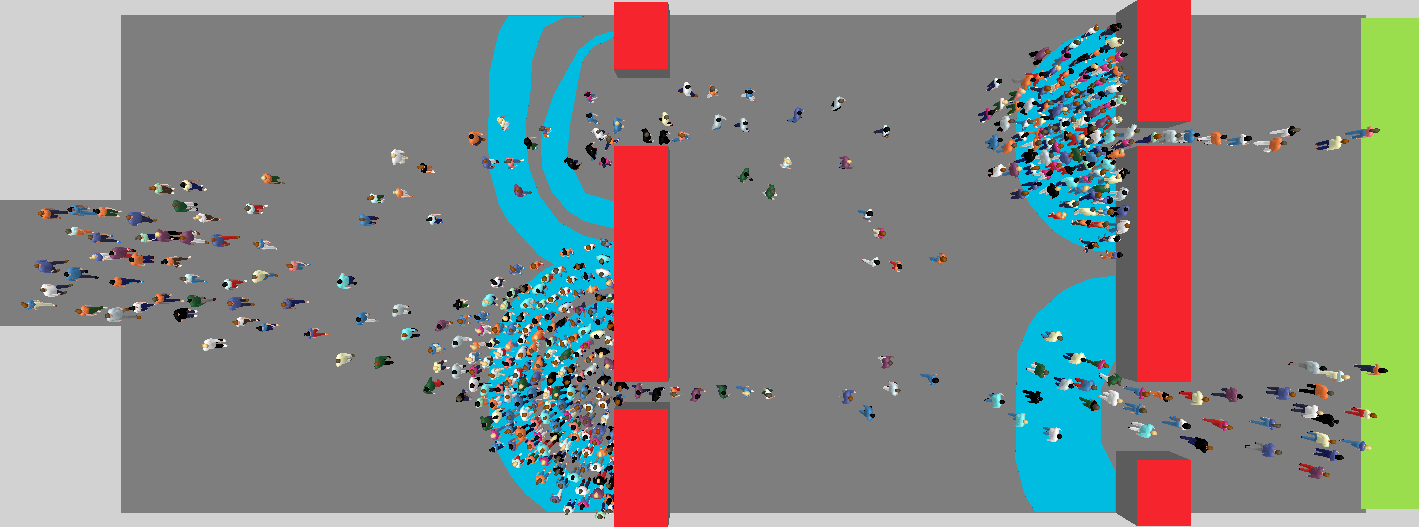} \\
\caption{Situation after 450 simulation seconds with a demand of 3 pedestrians per second: upper left: if pedestrians walk 50:50 the two shortest routes; upper right: if all pedestrians walk the route with highest capacity; lower left: if all routes are used equally; lower right: according to the result of the assignment process (39.4 : 0 : 17.2 : 43.4).}%
\label{fig:results3}%
\end{figure}

Figure \ref{fig:results3} shows a comparison of the situation after 450 simulation seconds as it occurs when different route choice ratios are chosen. One can get the impression -- and this is correct -- that pedestrians could be assigned more efficiently. The reason for this is that in our -- so far rather simple assignment method -- only pedestrians who arrive within the relevant time interval at the destination can contribute. In the discussion we will elaborate on how the calculation of route choice ratios could be improved.

Having looked to the details if demand is set to 3 pedestrians per second we will furthermore present the results of all demand volume variants we have considered. For this larger set of result data due to space limitations we have to restrict ourselves to choice ratios and travel times when equilibrium is reached or the assignment process has terminated otherwise. 

\begin{figure}[htbp]%
\center
\includegraphics[width=0.612\columnwidth]{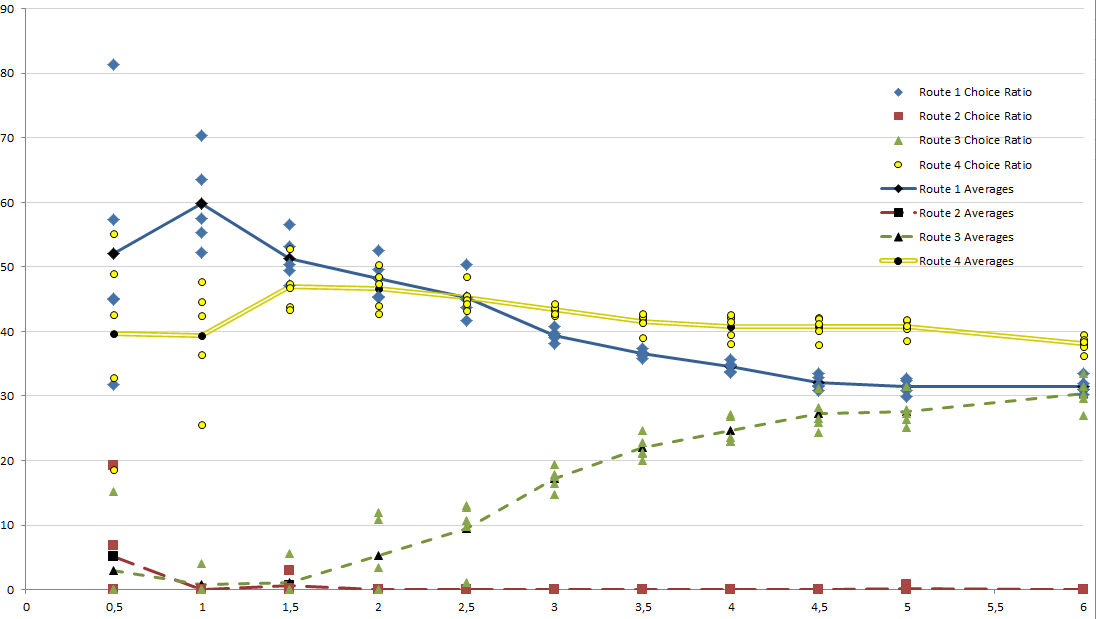} \\ \vspace{12pt}
\includegraphics[width=0.612\columnwidth]{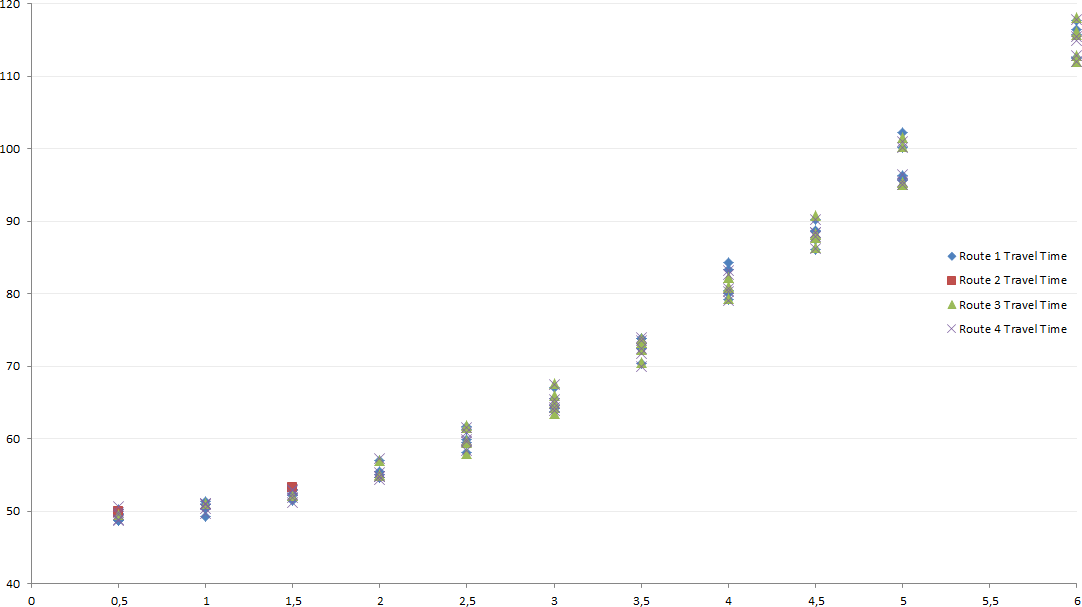}
\caption{Route choice ratios (top, in percent) and travel times (bottom, in seconds) vs. demand as they result from the assignment process. It can be seen how route 3 which leads through the two wider doors with increasing demand gains volume and thus leads to a more efficient walking behavior. It can also be seen how the travel times for a particular demand volume for all routes and all assignment processes all have a similar value compared to the overall dynamics of travel times with increasing demand. For small demand route 1 has a higher load than route 4 although both mainly differ in the sequence of bottleneck widths (route 1: wide, narrow; route 4: narrow, wide). Even the narrow door has a capacity to easily accomodate the demand (about half of the total demand) of the cases with 0.5 or 1 pedestrians per second total demand. Why then is there a difference between the two route choice ratios? The reason is that along route 4 in periods of above average demand pedestrians have to align in less time to pass through the narrow bottleneck. If pedestrians don't manage to do so some of them have a time delay and that is sufficient to trigger the observed differences.}%
\label{fig:results4}%
\end{figure}

Figure \ref{fig:results4} shows the route choice ratios and the travel times as they result from the assignment processes.

\section{Summary, Conclusions, and Outlook}
This study gave an example of how a recently proposed method to compute and model routing alternatives for pedestrians in simulations can be applied. The computation of routing alternatives is formulated such that the proposed alternatives are sufficiently different which means more precisely that the routes are separated by at least one obstacle (or non-walkable ground) of a certain, configurable minimum size. Pedestrians follow the routes by heading sequentially from intermediate destination to intermediate destination. Hereby they can prefer to walk into the direction of the shortest path -- which is the usual strategy in a model of operational pedestrian dynamics \cite{Schadschneider2009b} -- nevertheless as a consequence of the intermediate destinations globally they can walk arbitrary detours. The intermediate destinations are geometrically shaped such that no artificial sharp turns or artificial bottlenecks are introduced locally into the motion. In other words: if the intermediate destinations are not marked specifically an observer could not guess their presence from observing local pedestrian behavior.

The routes which result from this method can easily be used for an assignment computation to find the user equilibrium with regard to travel times (Wardrop's principle: ``No traveler can improve his or her travel time by choosing a different route.''). This was done in this contribution with a rather simple geometry which nevertheless allowed to see exemplified how the method of route and intermediate destination computation works.

Furthermore the method to compute from simulation results (travel times) the route choice ratios for the next iteration step was rather simple. It only relied on the simulation results of the very last iteration step. One can imagine that oscillations in the results could be reduced if a longer history of simulation results is considered. Second, the method ignores the current travel times of pedestrians who are still in the simulation when the relevant time interval ends. It is, however, a difference, if at that time all pedestrians who are still on their way have a current travel time below the one which has been measured as average for a particular route or if there are individuals who have had long delays and thus travel times well above that average. This is related to a third issue: we have considered all pedestrians who have arrived within a certain time interval. One could also try to consider pedestrians who depart within a certain time interval. This is more difficult as there is no guaranteed maximum simulation time, but after all the route decision occurs when a pedestrian departs, so naturally it makes sense to aggregate based on time intervals of departure.

Independent of the details of the assignment method it would be interesting to compare in future studies the results as well as the performance (mainly concerning computation time) of the proposed iterative method with one-shot methods where it is tried to achieve a good -- i.e. travel time-based near user equilibrium -- within one single simulation run \cite{Kretz2009c,Kretz2011e,Kretz2011f,kemloh2012modeling,Kretz2013c}.


\bibliographystyle{utphys2011}
\bibliography{PedAssignment}

\providecommand{\href}[2]{#2}\begingroup\raggedright\begin{thebibliography}{10}

\bibitem{wardrop1952some}
J.~Wardrop, ``Road Paper. Some theoretical aspects of road traffic research'',
  \href{http://dx.doi.org/10.1680/ipeds.1952.11259}{{\em {ICE} Proceedings:
  Engineering Divisions} {\bfseries 1} (1952) }.

\bibitem{Beckmann1956}
M.~Beckmann, C.~McGuire, and C.~Winsten, ``Studies in the Economics of
  Transportation'', tech. rep., Yale University Press, 1956.

\bibitem{leblanc1975efficient}
L.~LeBlanc, E.~Morlok, and W.~Pierskalla, ``An efficient approach to solving
  the road network equilibrium traffic assignment problem'',
  \href{http://dx.doi.org/10.1016/0041-1647(75)90030-1}{{\em Transportation
  Research} {\bfseries 9} no.~5, (1975) 309--318}.

\bibitem{bar2002origin}
H.~{Bar-Gera}, ``Origin-based algorithm for the traffic assignment problem'',
  \href{http://dx.doi.org/10.1287/trsc.36.4.398.549}{{\em Transportation
  Science} {\bfseries 36} no.~4, (2002) 398--417}.

\bibitem{gentile2009linear}
G.~Gentile and K.~Noekel, ``Linear User Cost Equilibrium: the new algorithm for
  traffic assignment in VISUM'', in {\em Proceedings of European Transport
  Conference 2009}.
\newblock 2009.

\bibitem{Kretz2014h}
T.~Kretz, K.~Lehmann, and I.~Hofs{\"a}{\ss}, ``User Equilibrium Route
  Assignment for Microscopic Pedestrian Simulation'', {\em submitted} ,
  \href{http://arxiv.org/abs/1401.0799}{{\ttfamily arXiv:1401.0799
  [physics.soc-ph]}}.

\bibitem{Kretz2013b}
T.~Kretz, K.~Lehmann, and T.~Friderich, ``Selected Applications of a Dynamic
  Assignment Method for Microscopic Simulation of Pedestrians'', in {\em
  European Transport Conference 2013}, p.~online.
\newblock 2013.

\bibitem{Kretz2014a}
T.~Kretz, K.~Lehmann, I.~Hofs{\"a}{\ss}, and A.~Leonhardt, ``Dynamic Assignment
  in Microsimulations of Pedestrians'', in {\em 93rd Annual Meeting of the
  Transportation Research Board}, p.~on CD.
\newblock 2014.
\newblock \href{http://arxiv.org/abs/1401.1308}{{\ttfamily arXiv:1401.1308
  [cs.CE]}}.

\bibitem{Heuvel2014}
J.~{van den Heuvel}, A.~Voskamp, W.~Daamen, and S.~Hoogendoorn, ``Using
  Bluetooth to estimate the impact of congestion on pedestrian route choice in
  train stations'', in {\em Traffic and Granular Flow 2013}, M.~Boltes,
  M.~Chraibi, A.~Schadschneider, and A.~Seyfried, eds., p.~to be published.
\newblock 2014.

\bibitem{vanwageningenkessels2014}
F.~{van Wageningen-Kessels}, W.~Daamen, and S.~Hoogendoorn, ``Pedestrian
  Evacuation Optimization -- Dynamic Programming in Continuous Space and
  Time'', in {\em Traffic and Granular Flow 2013}, M.~Boltes, M.~Chraibi,
  A.~Schadschneider, A.~Seyfried, and M.~Schreckenberg, eds.
\newblock 2014.

\bibitem{Viswalk540}
{PTV AG}, {\em {PTV Vissim 5.40 -- User Manual}}.
\newblock {PTV Group}, Haid-und-Neu-Str. 15, D-76131 Karlsruhe, Germany, 2012.
\newblock Version 5.40-08.

\bibitem{Kretz2008b}
T.~Kretz, S.~Hengst, and P.~Vortisch, ``{Pedestrian Flow at Bottlenecks --
  Validation and Calibration of VISSIM's Social Force Model of Pedestrian
  Traffic and its Empirical Foundations}'', in {\em {International Symposium of
  Transport Simulation 2008 (ISTS08)}}, M.~Sarvi, ed., p.~electronic
  publication.
\newblock Monash University, Gold Coast, Australia, 2008.
\newblock \href{http://arxiv.org/abs/0805.1788}{{\ttfamily arXiv:0805.1788
  [cs.MA]}}.

\bibitem{fellendorf2010microscopic}
M.~Fellendorf and P.~Vortisch,
  \href{http://dx.doi.org/10.1007/978-1-4419-6142-6_2}{``Microscopic traffic
  flow simulator VISSIM'',} in {\em Fundamentals of Traffic Simulation},
  pp.~63--93.
\newblock Springer, 2010.

\bibitem{Johansson2007}
A.~Johansson, D.~Helbing, and P.~Shukla, ``{Specification of the Social Force
  Pedestrian Model by Evolutionary Adjustment to Video Tracking Data}'',
  \href{http://dx.doi.org/10.1142/S0219525907001355}{{\em Advances in Complex
  Systems} {\bfseries 10} no.~4, (2007) 271--288},
  \href{http://arxiv.org/abs/0810.4587}{{\ttfamily arXiv:0810.4587
  [physics.soc-ph]}}.

\bibitem{msc1238}
{International Maritime Organization}, ``MSC Circ 1238'', tech. rep., IMO,
  2007.

\bibitem{Schadschneider2009b}
A.~Schadschneider, H.~Kl{\"u}pfel, T.~Kretz, C.~Rogsch, and A.~Seyfried,
  \href{http://dx.doi.org/10.4018/978-1-60566-226-8.ch006}{``{Fundamentals of
  Pedestrian and Evacuation Dynamics}'',} in {\em {Multi-Agent Systems for
  Traffic and Transportation Engineering}}, A.~Bazzan and F.~Kl{\"u}gl, eds.,
  ch.~{VI}, pp.~124--154.
\newblock Information Science Reference, Hershey, PA, USA, 2009.
\newblock ISBN:978-1-60566-226-8.

\bibitem{Kretz2009c}
T.~Kretz, ``{The use of dynamic distance potential fields for pedestrian flow
  around corners}'', in {\em First International Conference on Evacuation
  Modeling and Management}.
\newblock TU Delft, 2009.
\newblock \href{http://arxiv.org/abs/0906.2667}{{\ttfamily arXiv:0906.2667
  [cs.MA]}}.

\bibitem{Kretz2011e}
T.~Kretz, A.~Gro{\ss}e, S.~Hengst, L.~Kautzsch, A.~Pohlmann, and P.~Vortisch,
  ``{Quickest Paths in Simulations of Pedestrians}'',
  \href{http://dx.doi.org/10.1142/S0219525911003281}{{\em Advances in Complex
  Systems} {\bfseries 14} (2011) 733},
  \href{http://arxiv.org/abs/1107.2004}{{\ttfamily arXiv:1107.2004
  [physics.soc-ph]}}.

\bibitem{Kretz2011f}
T.~Kretz, S.~Hengst, V.~Roca, A.~{P{\'e}rez Arias}, S.~Friedberger, and
  U.~Hanebeck,
  \href{http://dx.doi.org/10.1109/ICCVW.2011.6130239}{``{Calibrating Dynamic
  Pedestrian Route Choice with an Extended Range Telepresence System}'',} in
  {\em 2011 IEEE International Conference on Computer Vision Workshops},
  pp.~166--172.
\newblock 2011.
\newblock First IEEE Workshop on Modeling, Simulation and Visual Analysis of
  Large Crowds, 6-13 November 2011, Barcelona, Spain.

\bibitem{kemloh2012modeling}
A.~Kemloh~Wagoum, A.~Seyfried, and S.~Holl, ``Modeling The Dynamic Route Choice
  Of Pedestrians To Assess The Criticality Of Building Evacuation'',
  \href{http://dx.doi.org/10.1142/S0219525912500294}{{\em Advances in Complex
  Systems} {\bfseries 15} no.~07, (2012) },
  \href{http://arxiv.org/abs/1103.4080}{{\ttfamily arXiv:1103.4080 [cs.OH]}}.

\bibitem{Kretz2013c}
T.~Kretz, ``{The Effect of Integrating Travel Time}'', in {\em Pedestrian and
  Evacuation Dynamics 2012}, U.~Weidmann, U.~Kirsch, and M.~Schreckenberg, eds.
\newblock 2014.
\newblock \href{http://arxiv.org/abs/1204.5100}{{\ttfamily arXiv:1204.5100
  [physics.soc-ph]}}.
\newblock (in press).

\end{thebibliography}\endgroup

\end{document}